\RequirePackage{silence}
\documentclass[preprint,12pt]{elsarticle}

\usepackage{amssymb}
\usepackage{amsmath}
\usepackage{amsmath}
\usepackage[version=4]{mhchem}
\usepackage[utf8]{inputenc} 

\usepackage[table]{xcolor}% http://ctan.org/pkg/xcolor
\usepackage{graphicx}
\graphicspath{./images/}

\usepackage{natbib}

\journal{BioSystems}

\begin{document}

\begin{frontmatter}



\title{Information Transmission and Processing in G-Protein-Coupled-Receptor Complexes} %% Article title


\author{Roger D. Jones $^{a,b,c,e}$}
\affiliation[UNC]
{Department of Biology, University of North Carolina at Chapel Hill, Chapel Hill, North Carolina 27514, USA}
\affiliation[UNIVE]{Dipartimento di Scienze Molecolari e Nanosistemi, Universit\`a Ca' Foscari Venezia, 30123 Venezia, Italy}
\affiliation[ECLT]{European Centre for Living Technology (ECLT) Ca' Bottacin, 3911 Dorsoduro Calle Crosera, 30123 Venezia, Italy}

\author{Achille Giacometti$^{b,c}$}

\author{Alan M. Jones$^{a,d}$}
\affiliation[UNC]
{Department of Pharmacology, University of North Carolina at Chapel Hill, Chapel Hill, North Carolina 27514, USA}

\affiliation[email]
{Corresponding author: RogerDJonesPhD@gmail.com}

%% Abstract
\begin{abstract}
We present a general theoretical framework for molecular computation in biological systems and apply it to G-protein-coupled receptors (GPCRs), which serve as central regulators of cellular information processing. 
Despite their importance, the physical principles underlying GPCR switching remain incompletely understood.
Using nonequilibrium thermodynamics, we construct a model that  identifies the parameters governing receptor-state transitions.

The framework shows that the configuration of the switch  is governed by two factors: the ATP/GTP-driven chemical flux through the receptor complex and the free-energy difference between competing switch states. 
The model predicts that GPCRs can occupy three quasistable configurations, corresponding to “on,” “off,” and an intermediate state, each representing a local maximum in information transmission. 
Switch states can also be characterized by whether the switch supports a net chemical flux.
Active states support sustained chemical flux, whereas inactive states do not.

The model incorporates reciprocal conformation-fit changes between ligand and receptor.  As such,  the model predicts that phosphatase activity, represented as an effective energy barrier, primarily determines whether the switch occupies the “on” or “off” state, whereas kinase activity maintains flux without directly setting state occupancy.  These predictions based on experimental evidence point to new targets for drug design.   
Comparison with label-free impedance measurements supports the existence of multiple quasistable states that depend on ligand conformation.

Because the framework relies on general nonequilibrium principles rather than system-specific biochemistry, it extends naturally to other biological switching systems driven by chemical flux.
\end{abstract}



%% Keywords

\end{frontmatter}



\section{Introduction}

Biological systems display complex behaviors that arise from interactions across multiple levels of organization. Although this complexity has led some to question whether biology can be described by a compact set of theoretical principles \cite{forestiero2022historical};
concepts from thermodynamics, homeostasis \cite{bechtel2024situating}, information theory \cite{bajic2024information}, and complexity science \cite{hebert2024path} promise, however, to provide useful organizing frameworks. 
The purpose of this study is to develop a first-principles description of molecular computation that draws on these organizing ideas and to apply it to the important example of G protein coupled receptors (GPCRs), a major class of signaling proteins with central roles in physiology and pharmacology \cite{zhang2024g, koenig2017precision,kosorok2019precision,ginsburg2018precision}.
Ultimately, the goal for such a model is to identify drug targets in order to improve health outcomes of drug therapies. 

Cells process environmental information through networks of chemical reactions that act as molecular switches \cite{qian2007phosphorylation,latorraca2018molecular}. 
These switches operate far from thermo-chemical equilibrium and require continuous input of energy. GPCRs are a clear example since their activity depends on GTP hydrolysis in G proteins and ATP driven phosphorylation and dephosphorylation cycles on the receptor. Without these fluxes, GPCR signaling would relax to equilibrium and lose functional capacity. This energy dependence suggests that biological computation may be shaped by evolutionary pressures that favor efficient transmission of information under physical constraints.

Recent experimental work highlights the need for theoretical models that account for these nonequilibrium features. 
Assays have begun to probe GPCR dynamics using photoisomerizable ligands and label-free measurements of cellular responses \cite{wirth2023monitoring}. In this study, we analyze such measurements within a nonequilibrium thermodynamic framework that identifies two key control parameters: the chemical flux through the relevant reaction cycles and the free-energy difference between competing switch states. These parameters inform the ligand-receptor configuration and the amount of information that can be transmitted across the membrane.
We also have theoretical indications that an important control parameter for switching is the height of the energy barrier that regulates flux from the on state to the off state.

The model predicts that GPCR switches can occupy up to three quasistable states, an on state, an off state, and an intermediate state, rather than the two states assumed in most biochemical models. The switch state depends on both the fixed chemical structure of the ligand and its inducible conformational changes. Experiments using photoactive ligands support this interpretation.

The theoretical approach is based on statistical mechanics and constrained optimization \cite{jaynes1957information, jaynes1957information2} and extends previous work that applied information theory to biochemical networks \cite{jones14model,jones2023proposed,jones2024information}. It also generalizes standard mass action and Markov chain formulations, which assume constant reaction rates. This assumption does not hold for systems embedded within protein matrices that are continually modified by ligand binding and by ATP driven entropy flow. Under such conditions, reaction rates depend on the nonequilibrium environment, and the quasistable solutions appear in the theory.

This paper is structured so that the main text presents general ideas, applications, and key theoretical results. The detailed mathematical development is provided in the Appendix, where a more complete description of the theoretical framework is given.

\section{Methods}

\subsection{Schematic of the Experiment}
\begin{figure}[hbt!] %s state preferences regarding figure placement here
% use to correct figure counter if necessary
%\renewcommand{\thefigure}{2}
\begin{center}
\includegraphics[width=0.7\textwidth]{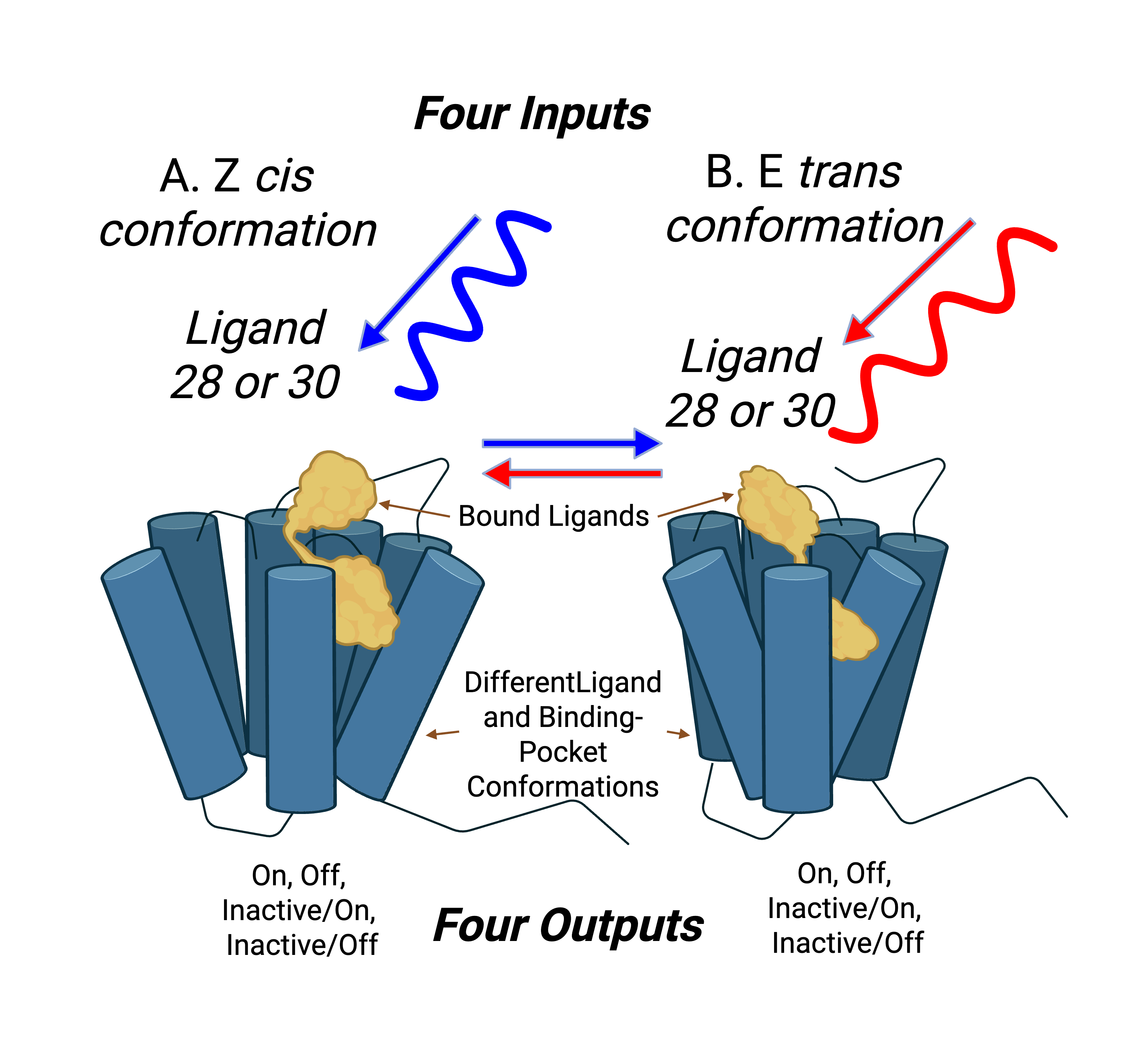}
\end{center}
\caption{ \textbf{Conceptual schematic of the experiment.}
Ligands in cis (Z) or trans (E) configurations bind to GPCRs, while electrical impedance across a cell layer monitors signaling activity. Two engineered ligands (28 and 30) reversibly switch between Z and E isomers under wavelength-specific light (blue: Z→E; red: E→Z). Each conformation induces a distinct GPCR state, yielding four input combinations, encoding up to two bits of information. 
}
\label{ExpTimeReversal2} % \label works only AFTER \caption within figure environment
\end{figure}

Conceptual schematic of the experimental setup from Wirth et al. \cite{wirth2023monitoring} is shown in Figure~\ref{ExpTimeReversal2}. 
Chinese hamster ovary cells (CHO) were genetically engineered to overexpress the receptor, a member of the neuropeptide Y family, and cultured to confluence on a gold foil substrate \cite{cabrele2000molecular}. This uniform expression of the receptor enabled for consistent detection of ligand-induced conformational changes via electronic impedance measurements across the gold foil.

Two wavelength-driven ligand isomers, cis (Z) and trans (E), were derived from azobenzene (ligand 28) and arylazopyrazole (ligand 30), producing four distinct GPCR inputs: E-28, Z-28, E-30, and Z-30. 
Z isomers were generated by 340 nm irradiation, while E-isomers were produced using 455 nm (ligand 28) and 528 nm (ligand 30) light. GPCR response was inferred from changes in cell morphology, monitored by impedance measurements \cite{skiba2022label,stolwijk2019increasing,wegener1999use}.

Natural downstream responses to activation of the 
$Y_4$
 NPY receptor include reduced cAMP levels, context-dependent activation of PLC, and engagement of MAPK signaling pathways. The impedance measurement can also be viewed as a downstream, label-free response. In this interpretation, the switch behaves as a single conglomerate system that reflects the combined activity of the GTPase switch in the G protein and the phosphorylation–dephosphorylation cycles that generate the barcode.

\subsubsection{Non-Mathematical Schematic of the Theory}

The mathematical details of the framework are presented in the Appendix. Here we provide a conceptual outline of the approach without relying on formal derivations.

Biological information processing requires molecular systems to operate persistently far from equilibrium. Cells maintain these non-equilibrium steady states through continuous energy flow from nucleotide hydrolysis, such as ATP or GTP consumption, with the resulting entropy dissipated into the surrounding thermal environment. These steady states are analogous to biological homeostasis, since they reflect stable yet energy-driven configurations.

The theoretical framework begins with Landauer’s Principle, which states that information is associated with a thermodynamic cost and is linked to heat and entropy production \cite{sagawa2018second}. When a molecular switch erases information, it dissipates heat into the thermal bath \cite{jones2025thermodynamics}. Because energy is conserved, creating or transmitting information requires work. In biochemical systems this work is provided by conformational changes in the protein matrix in which the switch resides. For GPCR signaling, this matrix includes the receptor and its associated G proteins. The GPCR–G protein complex acquires usable energy from the excess concentration of extracellular ligand, and the binding reaction that includes ligand, receptor, and G protein is a ternary process \cite{burger2024positive}.

The goal of the theory is to characterize the non-equilibrium steady states rather than the transient dynamics that lead to them. At steady state, the rate of information change is zero, which implies that the system occupies a local or global extremum of information flow. We assume that natural selection favors molecular organizations that maximize information flow under appropriate physical constraints. These constraints include the number of switches, the average energy of the switch configurations, and the average energy and entropy flux per switch.

Within this framework, optimization yields the fraction of time that each switch resides in inactive or active configurations, which correspond to states with zero or nonzero chemical flux, respectively. The theory also determines whether an active configuration is more likely to occupy the on or off state. These quantities provide the basis for predicting downstream cellular responses to ligand stimulation.

\subsection{Application to G Protein-Coupled Receptor Complexes}

To demonstrate and validate the proposed theoretical framework, we focus on the fundamental process of information processing on the cell membrane. Among such processes, signaling through GPCRs plays a central role, representing one of the most critical mechanisms in both biological regulation and medical intervention.

\begin{figure}[hbt!] %s state preferences regarding figure placement here
% use to correct figure counter if necessary
%\renewcommand{\thefigure}{2}
\centering
\includegraphics[width=0.6\textwidth]{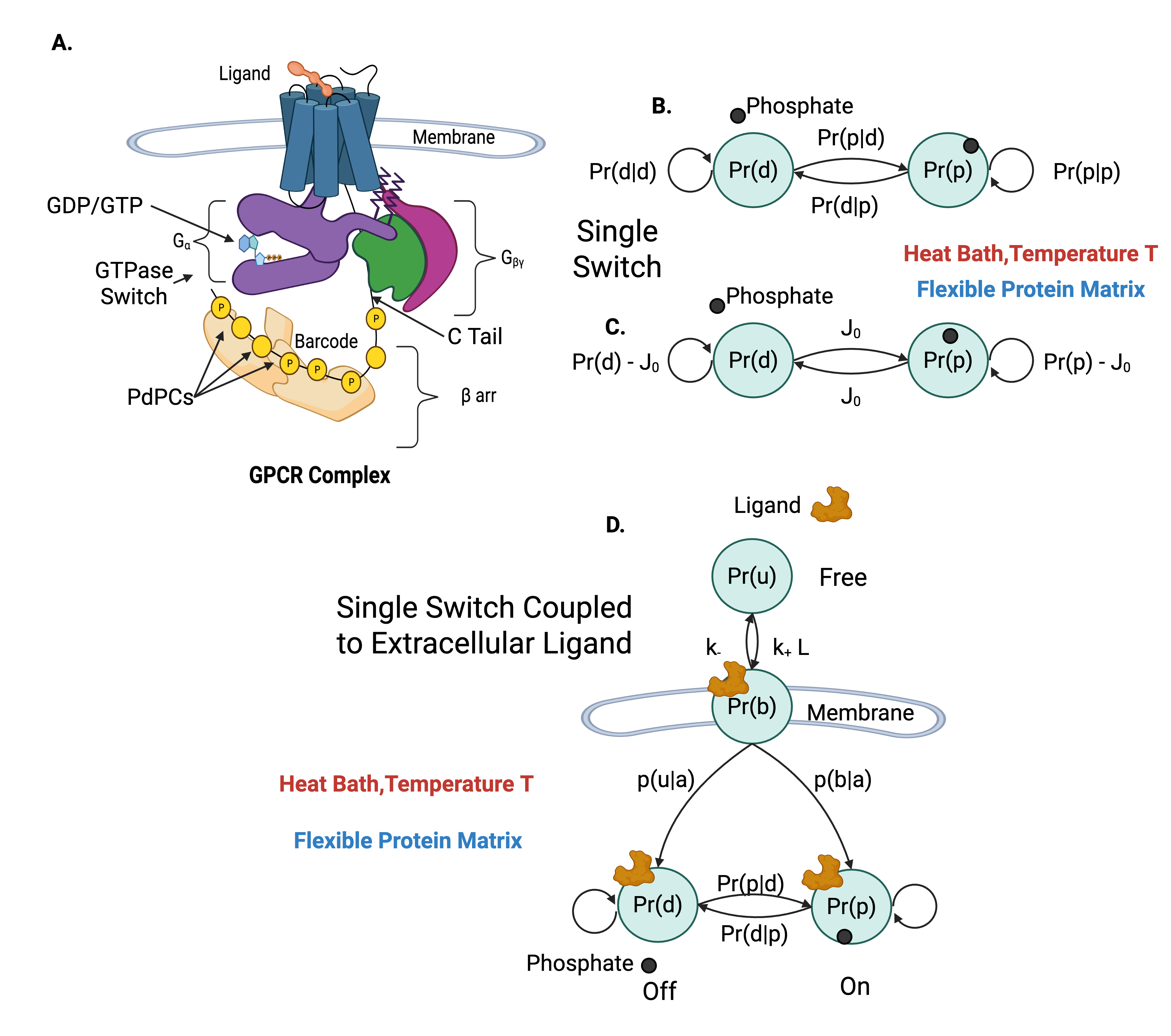}
\caption{ \textbf{Simple Molecular Switch}
{\bf A.}
The 7-transmembrane GPCR is illustrated with blue cylinders representing the seven $\alpha$ helices that span the cell membrane.
The extracellular ligand (orange) binds to the binding site of the GPCR inducing movement with the $\alpha$ helical bundles.
The helices alter the conformation of the intracellular loops (thin black lines) of the GPCR complex.
The intracellular portion of the complex has been separated for visibility.
Two pathways may be activated, the $G_{\alpha}$ pathway (purple) and the $\beta$arr pathway (tan).
The $G_{\alpha}$ subunit is a part of the G protein also composed of subunits $\beta$ and $\gamma$.
The $\beta$arr pathway is composed of additional response pathways determined by phosphorylation sites on the C tail of the GPCR and intracellular loops that form a barcode that encodes signals for downstream processes.
{\bf B.}
Picture of a single switch taken to be a PdPC.
A GTPase switch operates in the same manner.
The switch resides in a heat bath at temperature $T$.
In addition to the thermal bath, the switch is a component of a flexible protein complex that can modify energy barriers in the switch.
Here, $\Pr(d)$ and $\Pr(p)$ are the probabilities of finding the receptor to be dephosphorylated and phosphorylated, respectively, while $\Pr(p|d)$ and $\Pr(d|p)$ are the conditional probabilities of finding the receptor has transitioned between phosphorylated to dephosphorylated positions and {\it vice versa}.
{\bf C.}
Here, $J_0= \Pr(p|d)\Pr(d) = \Pr(d|p)\Pr(p)$ is the steady-state probability flux among states.
The flux is kept finite due to an external energy/entropy source.
{\bf D.}
Three-state model: free receptor (f), phosphorylated (p), and dephosphorylated (d). 
The bound state $b$ is bound to the ligand but not the phosphate.
The associated state $b$ has two internal states $d$ and $p$.
These two states form the $G_{\alpha}$ switch.
Ligand dissociation and association reaction rates are given by $k_-$ and $k_+$, respectively, and $L$ is the ligand concentration.
The picture for a GTPase switch is similar.
A G protein-GDP is bound to $\Pr(a)$ and the ligand through a ternary reaction. 
The on state occurs when GDP is replaced with GTP.
}
\label{SimpleSwitchBody} % \label works only AFTER \caption within figure environment
\end{figure}

The foundation of the current GPCR model is illustrated in Figure \ref{SimpleSwitchBody}, which schematizes the GPCR complex. In Figure~\ref{SimpleSwitchBody}A, the 7-transmembrane GPCR (blue) spans the membrane, interacting with both extracellular ligands (orange) and intracellular G proteins (purple, green, violet).

Ligand binding induces a conformational change in the GPCR, which enhances its affinity for heterotrimeric G proteins. This results in the formation of a ternary complex, comprising the ligand, GPCR, and G protein - that is more stable than the ligand receptor complex alone.\cite{mahoney2016mechanistic}.

The G protein $\alpha$ subunit ($G_{\alpha}$) initially binds to GDP, representing the off state of the GTPase switch \cite{qian2007phosphorylation}. Upon GPCR activation, GDP is released and replaced by GTP, whose cellular concentration is maintained far from equilibrium. This GTP binding activates $G_{\alpha}$, which completes the switch to the "on" state. The energy driving this transition derives from the thermodynamic disequilibrium between GDP and GTP.

The C-terminal tail of GPCR and the intracellular loops contain serine and threonine amino acid sites that can be phosphorylated/dephosphorylated by specific kinases/phosphatases. The C-terminal tail of GPCR and the intracellular loops contain serine and threonine amino acid sites that can be phosphorylated/dephosphorylated by specific kinases/phosphatases. 
The phosphorylation and dephosphorylation of these sites is known as a phosphorylation/dephosphorylation cycle (PdPC) that also acts as a molecular switch \cite{qian2007phosphorylation}. Similarly to the GTPase switch, the ``on" and ``off" states of these PdPC switches are determined by the phosphorylation status of specific sites. However, in this case, the switches are powered by the high concentration of adenosine triphosphate (ATP), which drives the process \cite{berg2002biochemistry}.

The collection of these PdPCs establishes a distinct molecular pattern, called a barcode, which serves as a guide for downstream signaling processes \cite{yang2017phosphorylation,chen2022qr}. 
Adapters read this phosphorylation barcode to propagate a certain response (Figure \ref{SimpleSwitchBody}). 
In animals, this adapter is called $\beta$-arrestin \cite{latorraca2020gpcr} and in plants it is called VPS26A/B \cite{jones2023vps26}. 

Furthermore, the flexible protein matrix surrounding the GPCR complex provides a mechanism to adjust energy levels and transition probabilities, facilitating changes in the conformation and behavior of the complex.

\section{Results}

\subsection{Experimental results relevant to this study}

If the four input conditions produce four independent output responses, this corresponds to two bits ($\log_2 4$) of information transmitted across the membrane. Fewer distinct outputs would indicate reduced information transfer. If only the chemical structure of the ligand (for example, 28 versus 30) influences the response, then only one bit of information is transmitted. Observation of more than two outputs therefore implies that both the fixed chemical structure and the induced-fit conformations of the ligand \cite{franco2021old} contribute to signaling.

The experiment reported in \cite{wirth2023monitoring} revealed four distinct responses, demonstrating that both the Z and E photoisomerized conformations, along with the fixed structural differences between ligands 28 and 30, play essential roles in determining receptor output. This result indicates that each ligand can exist in two functionally distinct conformational states after light activation, enabling the full transmission of two bits of information through the receptor complex. These findings are consistent with independent observations of GTPase activity in other G proteins that display similar two-bit information transmission \cite{keshelava2018high}.

The ligand conformation was found to switch reversibly between {\it cis} and {\it trans} forms even while the ligand remained bound to the receptor. Importantly, removal of extracellular ligand did not alter the downstream response, which indicates that signaling is determined by the conformation of the bound ligand rather than by its continued presence in solution.

\subsection{Theoretical Results}
\label{TheoreticalResults}

The model described in Sections \ref{TheoryAppendix}, \ref{Theoretical Model of the Switch} shows that a molecular switch can occupy quasistable states determined by two parameters: the chemical flux within the GPCR complex and the change of free energy of the switch. 
Both parameters are influenced by the characteristics of the extracellular ligand, its fixed and inducible conformations, and its concentration outside the cell.

This framework leads to the Biological Ensemble, a generalization of the Canonical Ensemble from equilibrium statistical mechanics \cite{reif2009fundamentals}.
The Canonical Ensemble, described by Feynman as the “summit of statistical mechanics”\footnote{``This fundamental law is the summit of statistical mechanics, and the entire subject is either the slide-down from this summit, as the principle is applied to various cases, or the climb-up to where the fundamental law is derived and the concepts of thermal equilibrium and temperature T clarified." R. P. Feynman} \cite[Chapter 1]{feynman2018statistical}, admits a single solution proportional to the exponential of the system energy \cite{reif2009fundamentals}. In contrast, the Biological Ensemble introduced here yields three possible solutions that turns the ensemble into a system of switches, a key result of this work.

We illustrate the power of the Biological Ensemble with an intuitive analogy. Near thermochemical equilibrium, a system relaxes to a single state, much as air molecules uniformly fill a room at fixed temperature and density. In our framework, if those molecules were in an NESS, the room could instead exist in one of three distinct states: one resembling the usual equilibrium condition and two others in which all molecules cluster into two different corners of the room. In this analogy, the NESS behaves as a switch with three possible configurations. It is clear that a room in equilibrium and a room in NESS are fundamentally different, and each would exhibit markedly different behavior.

\begin{figure}[hbt!] %s state preferences regarding figure placement here
% use to correct figure counter if necessary
%\renewcommand{\thefigure}{2}
\begin{center}
\includegraphics[width=1.0\textwidth]{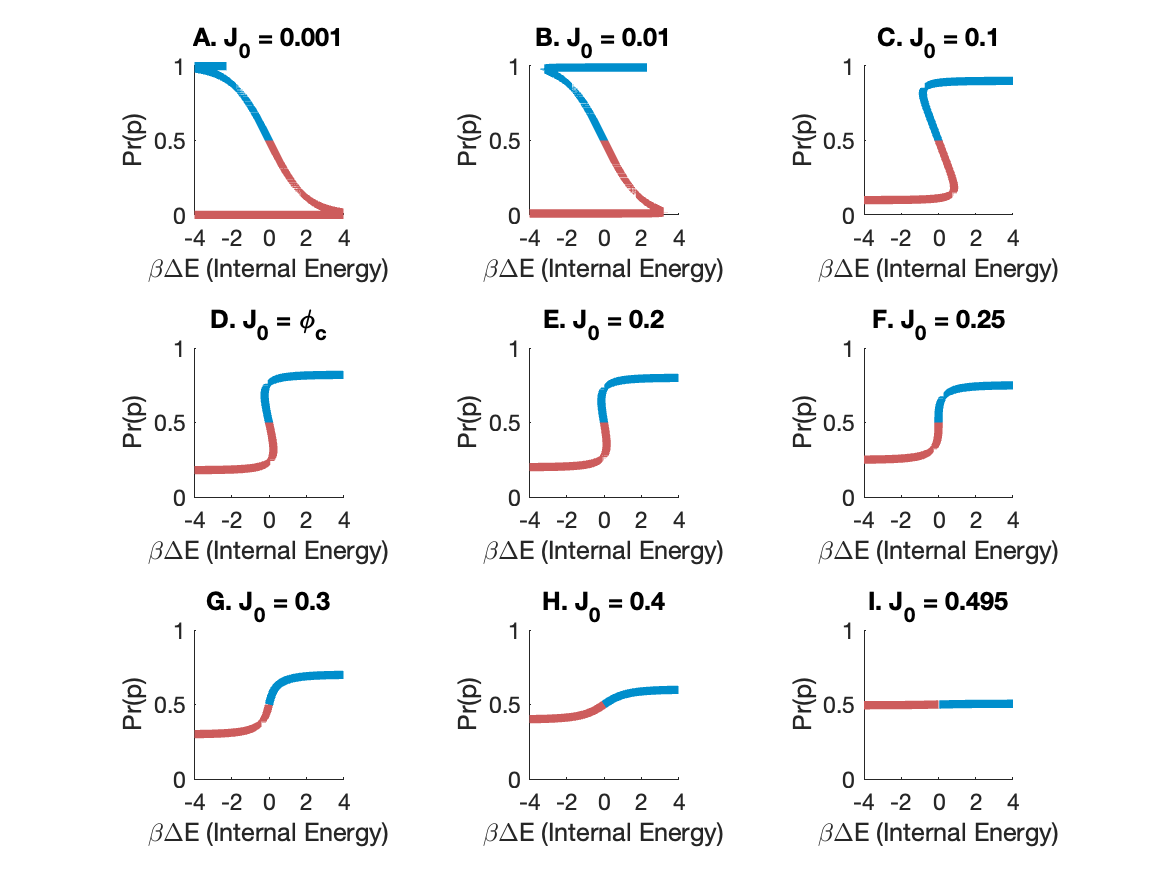}
\end{center}
\caption{
The quasistable solutions for $\Pr(p)$ for the Biological Ensemble as a function of the chemical flux $J_0$ and change in free energy between two receptor states $\beta \Delta E =\beta (E_p-E_d)$.
{\bf A.}
In the limit of very small flux, the solutions are (1) $\Pr(p) = 1,\; \Pr(d) = 1-\Pr(p)=0$ (blue), (2) $\Pr(d) =(1-\Pr(p)= 1,\; \Pr(p)=0$ (red), and (3) the intermediate solution ($0<\Pr(p)<1$).
{\bf B.-F.}
The  intermediate branch steepens as the chemical flux $J_0$ increases.
{\bf G.-I.}
For high values of flux $J_0$, the intermediate branch becomes less steep.
For fluxes less that the critical flux $ \phi_c \approx 0.182$
}
\label{SwitchQuasi} % \label works only AFTER \caption within figure environment
\end{figure}

The solutions correspond to local maxima of information flow, quasi-stable states that can transition under perturbation. Two parameters govern which state the switch occupies: (1) the energies of the phosphorylated ($E_p$) and dephosphorylated ($E_d$) conformations, and (2) the chemical flux $J_0$ from the phosphorylation/dephosphorylation cycle. Together, these parameters determine whether the switch is in the ``off", ``on" or intermediate state.

The energy parameter is given by the normalized change in the Gibbs free energy going from the ``off" to ``on" switch configuration.
\begin{equation} \label{energyParamter}
    \beta (E_p-E_d)
\end{equation}
where $\beta$ is the inverse of the temperature of the heat bath (Boltzmann constant = 1).
The Gibbs free energy is the valid form of free energy to use since the system is in a heat bath at temperature $T=1/\beta$ and no external work is being applied to the system.
The energies $E_p$ and $E_d$ are modified from their baseline measured values ($\approx 7$ kcal/mol) \cite{rosing1972value} by the presence of the protein matrix of the GPCR complex.
Modifications as high as $17$ kcal/mol have been seen in the muscles of athletes \cite{wackerhage1998recovery}.

If $R_p$ and $R_d$ are the numbers of receptors in the phosphorylated (``on") and dephosphorylated (``off") switch configurations, then the probability of the receptor being in each of the switch configurations is 
\begin{equation}
\Pr(p) = \frac{R_p}{R_d+R_p}
\end{equation}
and
\begin{equation}
    \Pr(d) = 1 -\Pr(p) = \frac{R_d}{R_d+R_p}
\end{equation}
for the phosphorylated and dephosphorylated configurations, respectively.

We define the chemical flux $J_0$ as the flow of probability from one switch configuration to the other.
\begin{equation} \label{fluxdef}
    J_0 = \Pr(d)\; \Pr(p|d) = \Pr(p)\; \Pr(d|p)
\end{equation}
where $\Pr(p|d)$ is the probability that the receptor state transitions from configuration $d$ to $p$ and $\Pr(d|p)$ is the probability of transition in the opposite direction.
The equality sign is a consequence of the steady-state requirement and is an expression of Bayes Theorem \cite{grover2012literature}.

The terms in Equation \ref{fluxdef} are intuitive from a chemist's perspective.
The probability $\Pr(d)$ is proportional to the chemical concentration of the system in the dephosphorylated configuration, while $\Pr(p)$ is proportional to the concentration of the phosphorylated concentration.
The transition probabilities $\Pr(p|d)$ and $\Pr(d|p)$ are the reaction rates.

An intuitive schematic of the information landscape is provided in \ref{LandscapeSchematic}, illustrating how the system navigates between quasistable states. 
The figure represents a landscape in which the curves are ridges of maximum switch information.
Three solutions exist for the probability $\Pr(p)$ that the switch is in the ``on" configuration for each flux and change in free energy value. 
In the limit of small flux, the solutions are (1) $\Pr(p) = 1,\; \Pr(d) = 1-\Pr(p)=0$ (blue), (2) $\Pr(d) =1-\Pr(p)= 1,\; \Pr(p)=0$ (red), and (3) the Canonical Ensemble (yellow).
We define solution (3) as the thermodynamic branch, while solutions (1) and (2) are defined to be the kinetic branch.

For small values of flux $J_0$, $\Pr(p)$ obeys the Canonical Ensemble.
The thermodynamic branch steepens as the chemical flux $J_0$ increases.
At $J_0 = \phi_c \approx 0.182$, the information content of the kinetic branch becomes equal to the information of the thermodynamic branch.
For $J_c > \phi_c \approx 0.182$, the information in the kinetic branch is greater than the information in the Thermodynamic Branch.
This implies that for flux greater than $\phi_c$, the states in the kinetic b ranch are more stable than the states in the thermodynamic branch.
The opposite is true for $J_0 < \phi_c$.

At $J_0 = 1/4$, the thermodynamic branch is given by $\Pr(p) =0$.
For $J_0 > 1/4$,
a discontinuity forms in the kinetic branch and the two solutions exchange roles as the change in free energy increases.

The curves in Figure \ref{SwitchQuasi} represent local maxima of information transmission, corresponding to quasi-stable states defined by chemical flux $J_0$
and change in free energy $\beta (E_p-E_d)$. 
Here $\beta$ is the inverse temperature of the heat bath, $E_p$ is the energy of the phosphorylated ``on" state and $E_d$ is the energy of the dephosphorylated ``off" state.
These configurations act as local attractors, indicating preferred switch states within the control space.

Observed switch behavior is expected to align with these quasi-stable states. The Biological Ensemble framework shows how variations in 
$J_0$
  and 
$\beta (E_p-E_d)$
guide the system into specific configurations, offering predictive insight into the switch dynamics under varying conditions.

A simple water-pump analogy, presented in the Discussion, offers a macroscopic visualization of the mechanism by which the switch transitions between these states.

\begin{figure}[hbt!] %s state preferences regarding figure placement here
% use to correct figure counter if necessary
%\renewcommand{\thefigure}{2}
\centering
\includegraphics[width=1.0\textwidth]{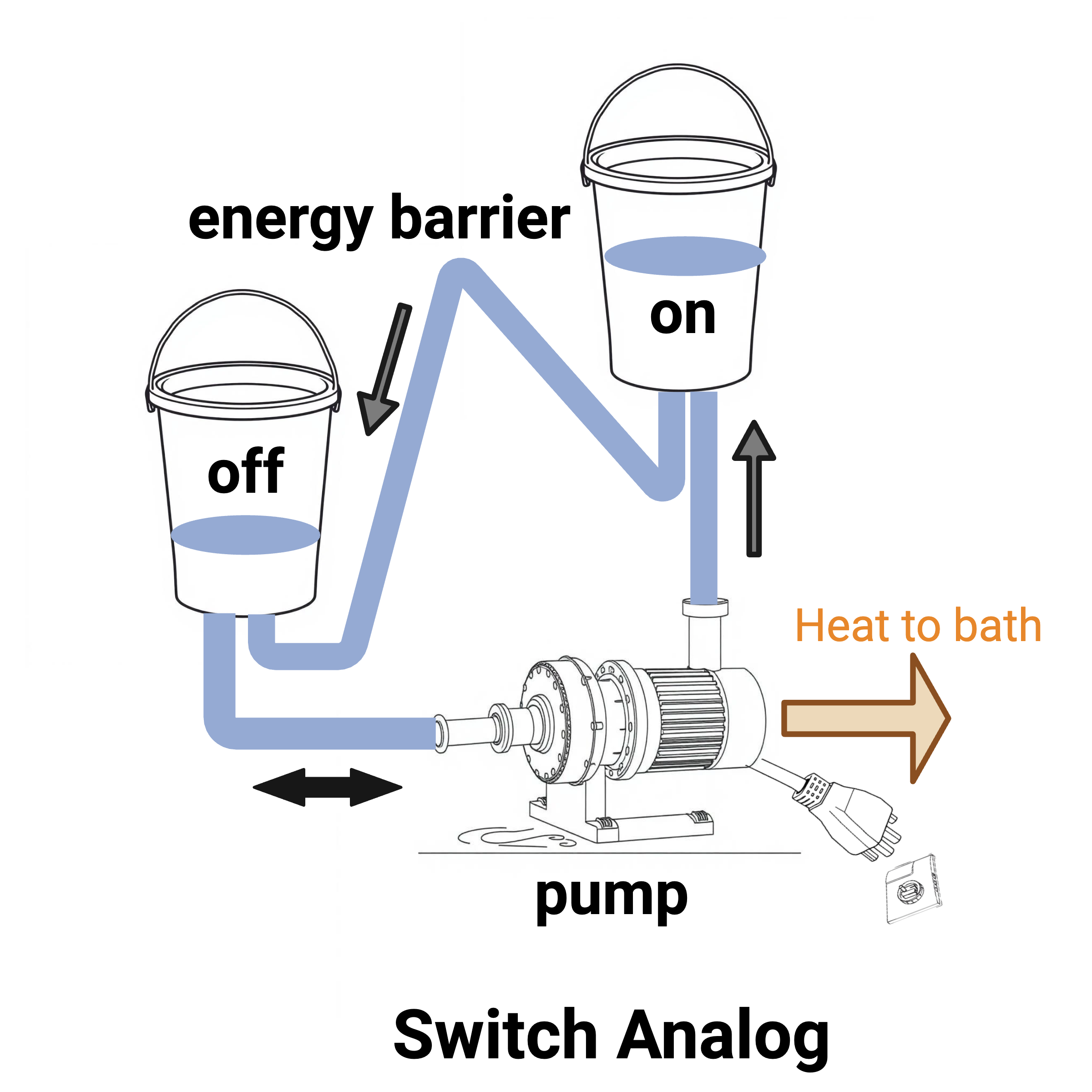}
\caption{ 
The “on” and “off” switch conformations are modeled as water buckets labeled on and off, where water levels represent the probability of each receptor state. Forward flow fills the on bucket, while reverse flow fills the off bucket. A pump, analogous to ATP/GTP-driven energy input, powers the cycle. The flow magnitude and state occupancy are regulated by an energy barrier, represented by the height of the return tube, which controls resistance to reverse flow.
}
\label{WaterExperiment} % \label works only AFTER \caption within figure environment
\end{figure}

A conceptual analogy \cite{jones2025plumbing} is provided in Figure~\ref{WaterExperiment}.
A pump drives water between two buckets that represent the on and off receptor states, while water height reflects state probability. The pump, analogous to an ATP or GTP driven kinase, maintains chemical flux. The energy barrier, represented by the return tube height, models phosphatase activity that regulates resistance to reverse flow. Heat generated by the pump dissipates into the surrounding environment. In this analogy, the phosphatase determines relative occupancy of the on and off states, whereas the kinase maintains flux.

The macroscopic hydraulic analog has three solutions similar to the three solutions in the mesoscopic model in Figures \ref{SwitchQuasi}.

\subsection{Comparison of Theory and Experimental Results}\label{comparison}

\begin{figure}[hbt!] %s state preferences regarding figure placement here
% use to correct figure counter if necessary
%\renewcommand{\thefigure}{2}
\begin{center}
\includegraphics[width=1.0\textwidth]{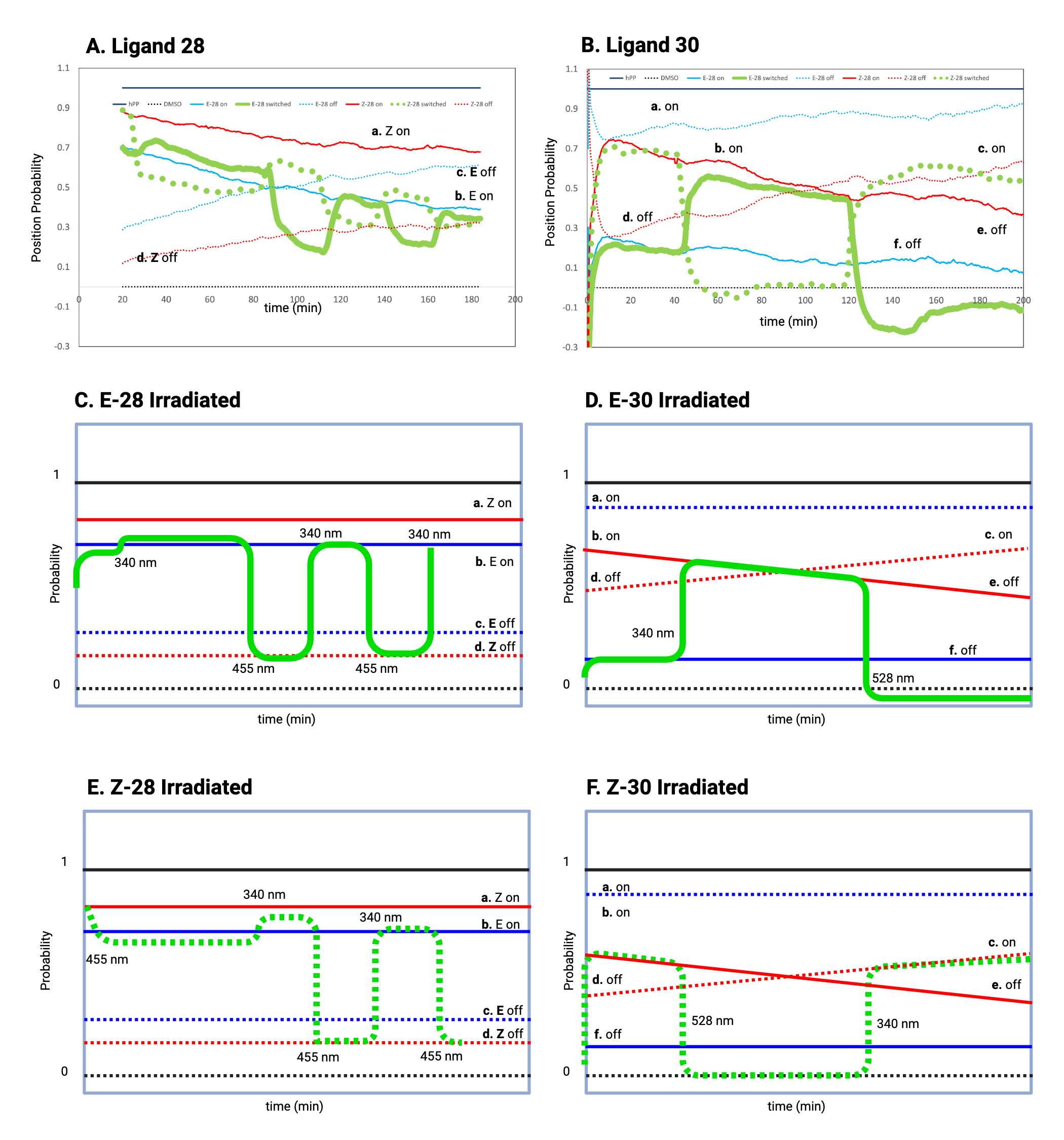}
\end{center}
\caption{ 
{\bf A. and B.} Experimental Results
{\bf C.-F.} Schematic of Quasistable States.
For Ligand 28, irradiation with wavelengths 340nm and 455 nm toggle the the system between the E/on (trans) and the Z/off (cis) states.
For Ligand 30, irradiation with 340 nm and 528 nm toggles between active and inactive configurations.
}
\label{Jumps} % \label works only AFTER \caption within figure environment
\end{figure}

The experimental results from Wirth et al. are depicted in Figures \ref{Jumps}A and B, alongside the theoretical predictions.
The mapping from impedance to probability is given in \ref{Connection Between Experiment and Theory}.
Solid and dashed red lines show impedance responses for ligands 28 and 30 initially prepared in the Z/on and Z/off configurations, respectively; blue lines represent E/on (solid) and E/off (dashed) preparations. The solid black line shows the response to the endogenous ligand hPP, and the dashed black line corresponds to the solvent control (DMSO). Figure \ref{Jumps}A shows the results for ligand 28, and Figure \ref{Jumps}B for ligand 30. Green lines indicate ligands prepared in the Z/on (dashed) and E/on (solid) states that were alternately irradiated with short (340 nm) and long (455 nm for 28; 528 nm for 30) wavelengths.

Figure \ref{Jumps}C–F present schematic summaries of the experimental transitions among quasi-stable switch configurations.
Figure \ref{Jumps}C shows that for ligand 28 in the E/on configuration, the impedance remains similar to that of Z/on. The off'' states (E/off and Z/off) also overlap. Initial 340 nm irradiation slightly increases impedance but does not shift the system out of E/on. Subsequent alternating irradiation causes the system to toggle between E/on and Z/off, indicating that cis (Z) favors the ``on'' state and trans (E) favors ``off.''

Figure \ref{Jumps}E shows the system initialized in the Z/on state. After a brief stable period, alternating light induces switching between Z/off and E/on, again confirming the correlation between the cis/trans states and switching on/off for ligand 28.

For ligand 30 (Figures~\ref{Jumps}D and F), the Z/on and Z/off responses overlap, indicating that the “on”/“off” distinction does not apply. Instead, the system transitions between an active Z state that supports chemical flux and an inactive E state that does not. Thus, ligand 30 modulates switch activity versus inactivity, rather than discrete “on”/“off” signaling.
Ligand 28 toggles the switch between on and off based on Z/E conformation, while ligand 30 toggles between active and inactive states. These behaviors reflect different modes of ligand control over GPCR switching.

\begin{figure}[hbt!] %s state preferences regarding figure placement here
% use to correct figure counter if necessary
%\renewcommand{\thefigure}{2}
\includegraphics[width=1.0\textwidth]{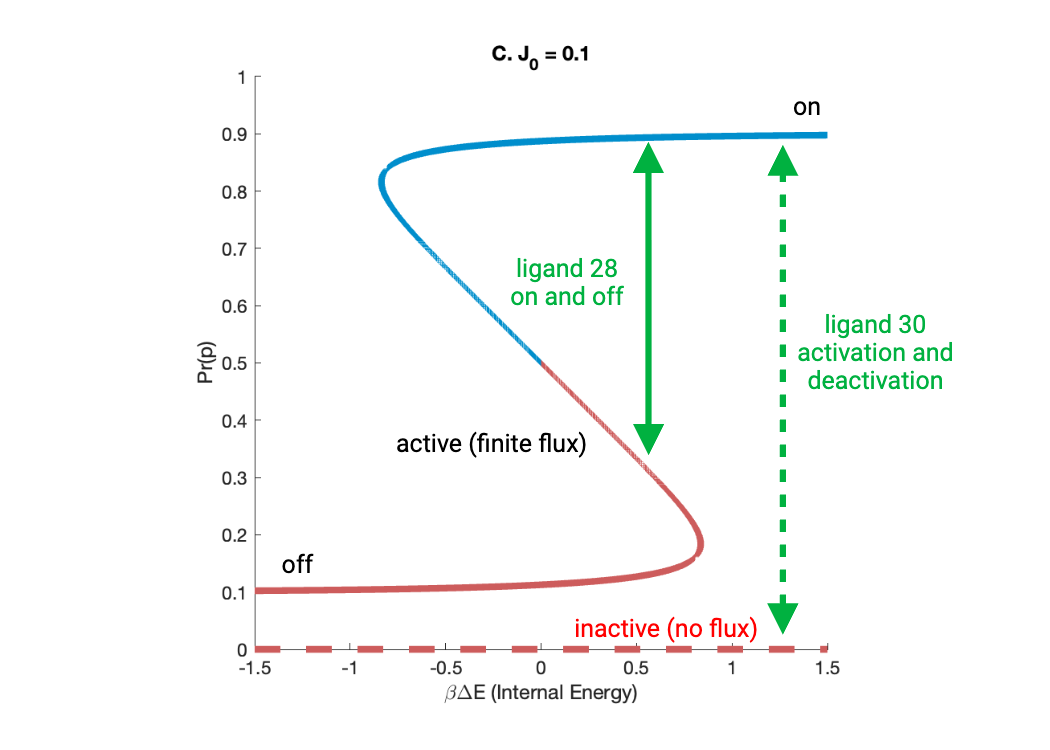}
\caption{
Enlargement of Figures \ref{SwitchQuasi}E and \ref{SwitchQuasi}A.
The solid blue and red curves are the active on ($\Pr(p)\ge 1/2$) and active off ($\Pr(p)< 1/2$)  states of the switch for $J_0=0.1$.
The dashed red curve is the off state for an inactive switch ($J_0\rightarrow 0$).
The theory predicts that ligand 28 in the Wirth experiment toggles between on and off states, depending on the ligand conformation, while ligand 30 toggles between the active on state and the inactive state of the switch.
}
\label{BlowUp} % \label works only AFTER \caption within figure environment
\end{figure}

The mapping of the experimental observations onto the theoretical framework is shown in Fig. \ref{BlowUp}, which enlarges the regions highlighted in Figs. \ref{SwitchQuasi}E and A. The theory assumes that the system preferentially occupies active portions of the solution curves that support high information flow, indicated in black. The inactive state, corresponding to zero phosphorylation, appears as a dashed red curve. Ligand 28 alternates between active on and active off states depending on its conformation, whereas ligand 30 toggles between an active on state and an inactive off state.

In Fig. \ref{BlowUp}, the flux was set to 
$J_0=0.1$. 
When 
$\beta (E_p-E_d)$ is much less than one, thermal fluctuations destabilize the quasistable states. When it is much greater than one, the states become so stable that switching is no longer feasible, preventing computation. This tradeoff was already recognized by Schrödinger in early discussions of biological information flow \cite{schrodinger1992life,phillips2021schrodinger}. The Wirth experiment therefore implies that the flux 
$J_0$ and the energy 
$\beta (E_p-E_d)$ fall near 0.1 and 1, respectively, illustrating how these otherwise inaccessible parameters may be inferred indirectly from experimental assays.

\section{Discussion}

Biological switches function within nonequilibrium environments where continuous energy flow is required to sustain activity. The results presented here illustrate how such switches occupy nonequilibrium steady states rather than the equilibrium states emphasized in classical thermodynamics. These steady states depend on both chemical flux and free-energy differences, and together they generate a richer set of molecular responses than can be obtained from equilibrium analysis alone. The model shows that molecular computation arises naturally from the balance among energy dissipation, information flow, and structural constraints imposed by the protein matrix in which the switch resides.

A central strength of the thermodynamic approach is its ability to identify system-level properties while avoiding the need for detailed microscopic descriptions. Many complex systems converge to characteristic final states that can be predicted from coarse variables. Molecular switches behave in the same way. Although the microscopic paths taken during signaling are highly heterogeneous, the steady states reached by the system depend primarily on the underlying energy landscape and on the fluxes that maintain the system away from equilibrium. This perspective aligns with Jaynes’s information-theoretic approach \cite{jaynes1957information}, which seeks to identify the macroscopic constraints that shape system behavior.

In the context of GPCRs, chemical flux provides the driving force that maintains switching among intermediate, off, and on configurations. Information erasure within these molecular machines produces heat, and the associated entropy flow requires sustained input of ATP or GTP. To capture this behavior, we introduced the Biological Ensemble, which extends classical free-energy minimization to include constraints from chemical flux and information transmission. This formulation predicts switching dynamics that are not accessible through equilibrium models or through kinetic schemes that assume constant reaction rates.

A hydraulic analogy helps illustrate these principles. In this analogy, the pump represents the kinase activity that sustains chemical flux, and the barrier height represents phosphatase activity that regulates transitions between states. Increasing the barrier shifts occupancy toward the on state, while lowering it favors the off state. If the barrier becomes too large, flux cannot be maintained, which reflects the loss of function when molecular systems become trapped in high-energy configurations. This analogy highlights the distinct roles of kinase and phosphatase activities in determining both flux and state occupancy.

The theory predicts two broad regimes of behavior. Near equilibrium, at low flux, the system follows thermodynamic predictions dominated by thermal fluctuations. As flux increases, the system enters a kinetic regime characterized by nonlinear effects and the appearance of multiple quasistable states. These additional states represent local maxima in information transmission and cannot be captured by traditional mass-action or Markov-chain models that assume fixed reaction rates. The transition between regimes marks a qualitative shift in system organization and information-processing capacity.

Ligand structure and availability regulate both the free-energy difference between states and the flux that sustains switching. The impedance measurements analyzed here reveal four distinct outputs that correspond to combinations of active and inactive states and on and off configurations. Because the measurements are label-free, the assay does not distinguish between GTPase-driven and phosphorylation-driven mechanisms, yet the theoretical framework applies equally to both. The presence of four outputs indicates that ligand conformation and ligand identity contribute independent bits of information, consistent with two-bit transmission in other G-protein systems.

The theory also identifies control parameters that are not considered in precursor models, including the barrier height that determines the ease of transition between states. Further controls are expected to emerge when the theory is extended to arrays of interacting switches, and this will be addressed in future work. 
The transition from inactive switch states ($J=0$) to active states ($J>0$) is subtle and requires the interaction of multiple switches.
This is the topic of follow-up work.
Additional experimental comparisons, particularly those that examine downstream branches of GPCR signaling, will be essential for evaluating the generality of the framework.

\section{Acknowledgements}
We thank Prof. Joachim Wegener and his team for generously providing the raw data used in our analysis. We also thank Katherine Newhall for valuable insights on macroscopic reversibility.
Research by AG is supported by MIUR PRIN-COFIN2022  grant 2022JWAF7Y.
The figures were generated with Blender. 
The cubic equations were solved with Mathematica.
Numerical calculations were performed with MATLAB.

\section{Author Contributions}
RDJ: Conceptualization, Formal analysis, Investigation, Methodology, Project administration, Software, Writing – original draft.
AMJ: Conceptualization, Investigation, Validation, Writing – review and editing.
AG: Conceptualization, Formal analysis, Methodology, Writing – review and editing.

\section{Funding}
Research in the laboratory of AJ is supported by the National Science Foundation (MCB-0718202) and the National Institute of General Medical Sciences (R01GM065989).


%\bibliographystyle{unsrtnat}

%\printbibliography


%
\bibliographystyle{splncs04}
\bibliography{library}

@article{burger2024positive,
  title={Positive allosteric modulation of a GPCR ternary complex},
  author={Burger, Wessel AC and Draper-Joyce, Christopher J and Valant, Celine and Christopoulos, Arthur and Thal, David M},
  journal={Science Advances},
  volume={10},
  number={37},
  pages={eadp7040},
  year={2024},
  publisher={American Association for the Advancement of Science}
}

@article{jones2025plumbing,
  title={Plumbing Analog of Molecular Computation},
  author={Jones, Roger D and Giacometti, Achille and Jones, Alan M},
  journal={arXiv preprint arXiv:2511.20339},
  year={2025}
}

@article{jones2025thermodynamics,
  title={Thermodynamics of Biological Switches},
  author={Jones, Roger D and Giacometti, Achille and Jones, Alan M},
  journal={arXiv preprint arXiv:2510.25359},
  year={2025}
}

@article{zhang2024g,
  title={G protein-coupled receptors (GPCRs): advances in structures, mechanisms and drug discovery},
  author={Zhang, Mingyang and Chen, Ting and Lu, Xun and Lan, Xiaobing and Chen, Ziqiang and Lu, Shaoyong},
  journal={Signal transduction and targeted therapy},
  volume={9},
  number={1},
  pages={88},
  year={2024},
  publisher={Nature Publishing Group UK London}
}

@article{bechtel2024situating,
  title={Situating homeostasis in organisms: maintaining organization through time},
  author={Bechtel, William and Bich, Leonardo},
  journal={The Journal of Physiology},
  volume={602},
  number={22},
  pages={6003--6020},
  year={2024},
  publisher={Wiley Online Library}
}

@article{bajic2024information,
  title={Information Theory, Living Systems, and Communication Engineering},
  author={Baji{\'c}, Dragana},
  journal={Entropy},
  volume={26},
  number={5},
  pages={430},
  year={2024},
  publisher={MDPI}
}

@article{hebert2024path,
  title={The path of complexity},
  author={H{\'e}bert-Dufresne, Laurent and Allard, Antoine and Garland, Joshua and Hobson, Elizabeth A and Zaman, Luis},
  journal={npj Complexity},
  volume={1},
  number={1},
  pages={4},
  year={2024},
  publisher={Nature Publishing Group UK London}
}

@article{forestiero2022historical,
  title={The historical nature of biological complexity and the ineffectiveness of the mathematical approach to it},
  author={Forestiero, Saverio},
  journal={Theory in Biosciences},
  volume={141},
  number={2},
  pages={213--231},
  year={2022},
  publisher={Springer}
}

@article{grover2012literature,
  title={A literature review of Bayes’ theorem and Bayesian belief networks (BBN)},
  author={Grover, Jeff},
  journal={Strategic economic decision-making: using Bayesian belief networks to solve complex problems},
  pages={11--27},
  year={2012},
  publisher={Springer}
}

@article{franco2021old,
  title={The old and new visions of biased agonism through the prism of adenosine receptor signaling and receptor/receptor and receptor/protein interactions},
  author={Franco, Rafael and Rivas-Santisteban, Rafael and Reyes-Resina, Irene and Navarro, Gemma},
  journal={Frontiers in Pharmacology},
  volume={11},
  pages={628601},
  year={2021},
  publisher={Frontiers Media SA}
}

@article{rosing1972value,
  title={The value of $\Delta$G° for the hydrolysis of ATP},
  author={Rosing, J and Slater, EC},
  journal={Biochimica et Biophysica Acta (BBA)-Bioenergetics},
  volume={267},
  number={2},
  pages={275--290},
  year={1972},
  publisher={Elsevier}
}

@article{wackerhage1998recovery,
  title={Recovery of free ADP, Pi, and free energy of ATP hydrolysis in human skeletal muscle},
  author={Wackerhage, Henning and Hoffmann, Uwe and Essfeld, Dieter and Leyk, Dieter and Mueller, Klaus and Zange, Jochen},
  journal={Journal of applied physiology},
  volume={85},
  number={6},
  pages={2140--2145},
  year={1998},
  publisher={American Physiological Society Bethesda, MD}
}

@incollection{sagawa2018second,
  title={Second law, entropy production, and reversibility in thermodynamics of information},
  author={Sagawa, Takahiro},
  booktitle={Energy Limits in Computation: A Review of Landauer’s Principle, Theory and Experiments},
  pages={101--139},
  year={2018},
  publisher={Springer}
}

@article{jaynes1957information,
  title={Information theory and statistical mechanics},
  author={Jaynes, Edwin T},
  journal={Physical review},
  volume={106},
  number={4},
  pages={620},
  year={1957},
  publisher={APS}
}

@article{jaynes1957information2,
  title={Information theory and statistical mechanics. II},
  author={Jaynes, Edwin T},
  journal={Physical review},
  volume={108},
  number={2},
  pages={171},
  year={1957},
  publisher={APS}
}

@article{cabrele2000molecular,
  title={Molecular characterization of the ligand--receptor interaction of the neuropeptide Y family},
  author={Cabrele, Chiara and Beck-Sickinger, Annette G},
  journal={Journal of peptide science: an official publication of the European Peptide Society},
  volume={6},
  number={3},
  pages={97--122},
  year={2000},
  publisher={Wiley Online Library}
}

@article{donertacs2011role,
  title={Role of thought experiments in solving conceptual physics problems},
  author={D{\"o}nerta{\c{s}}, {\c{S}}ule},
  year={2011},
  publisher={Middle East Technical University}
}

@article{ireson2005einstein,
  title={Einstein and the nature of thought experiments},
  author={Ireson, GP},
  journal={School Science Review},
  volume={86},
  number={317},
  pages={47--53},
  year={2005},
  publisher={Association for Science Education (ASE)}
}

@article{morowitz2007energy,
  title={Energy flow and the organization of life},
  author={Morowitz, Harold and Smith, Eric},
  journal={Complexity},
  volume={13},
  number={1},
  pages={51--59},
  year={2007},
  publisher={Wiley Online Library}
}

@article{mahoney2016mechanistic,
  title={Mechanistic insights into {GPCR--G} protein interactions},
  author={Mahoney, Jacob P and Sunahara, Roger K},
  journal={Current opinion in structural biology},
  volume={41},
  pages={247--254},
  year={2016},
  publisher={Elsevier}
}

@incollection{skiba2022label,
  title={Label-free impedance measurements to unravel biomolecular interactions involved in G protein-coupled receptor signaling},
  author={Skiba, Michael and Stolwijk, Judith A and Wegener, Joachim},
  booktitle={Methods in Cell Biology},
  volume={169},
  pages={221--236},
  year={2022},
  publisher={Elsevier}
}

@article{stolwijk2019increasing,
  title={Increasing the throughput of label-free cell assays to study the activation of G-protein-coupled receptors by using a serial agonist exposure protocol},
  author={Stolwijk, Judith A and Skiba, Michael and Kade, Christian and Bernhardt, G{\"u}nther and Buschauer, Armin and H{\"u}bner, Harald and Gmeiner, Peter and Wegener, Joachim},
  journal={Integrative Biology},
  volume={11},
  number={3},
  pages={99--108},
  year={2019},
  publisher={Oxford University Press}
}

@article{wegener1999use,
  title={Use of electrochemical impedance measurements to monitor $\beta$-adrenergic stimulation of bovine aortic endothelial cells},
  author={Wegener, Joachim and Zink, Sigrid and R{\"o}sen, Peter and Galla, H-J},
  journal={Pfl{\"u}gers Archiv},
  volume={437},
  pages={925--934},
  year={1999},
  publisher={Springer}
}

@article{wirth2023monitoring,
  title={Monitoring the Reversibility of GPCR Signaling by Combining Photochromic Ligands with Label-free Impedance Analysis},
  author={Wirth, Ulrike and Erl, Julia and Azzam, Saphia and H{\"o}ring, Carina and Skiba, Michael and Singh, Ritu and Hochmuth, Kathrin and Keller, Max and Wegener, Joachim and K{\"o}nig, Burkhard},
  journal={Angewandte Chemie},
  volume={135},
  number={21},
  pages={e202215547},
  year={2023},
  publisher={Wiley Online Library}
}

@article{koenig2017precision,
  title={What is precision medicine?},
  author={Koenig, Inke R and Fuchs, Oliver and Hansen, Gesine and von Mutius, Erika and Kopp, Matthias V},
  journal={European respiratory journal},
  volume={50},
  number={4},
  year={2017},
note = {doi: 10.1183/13993003.00391-2017. PMID: 29051268.},
  publisher={Eur Respiratory Soc}
}

@article{kosorok2019precision,
  title={Precision medicine},
  author={Kosorok, Michael R and Laber, Eric B},
  journal={Annual review of statistics and its application},
  volume={6},
  pages={263--286},
  year={2019},
note = {doi: 10.1146/annurev-statistics-030718-105251. PMID: 31073534; PMCID: PMC6502478.},
  publisher={Annual Reviews}
}

@article{ginsburg2018precision,
  title={Precision medicine: from science to value},
  author={Ginsburg, Geoffrey S and Phillips, Kathryn A},
  journal={Health Affairs},
  volume={37},
  number={5},
  pages={694--701},
note = {doi: 10.1377/hlthaff.2017.1624. PMID: 29733705; PMCID: PMC5989714.},
  year={2018}
}

@book{ash2012information,
  title={Information theory},
  author={Ash, Robert B},
  year={2012},
  publisher={Courier Corporation}
}

@book{feynman2018statistical,
  title={Statistical mechanics: a set of lectures},
  author={Feynman, Richard P},
  year={2018},
  publisher={CRC press}
}

@article{jones14model,
  title={Model of Ligand-Triggered Information Transmission in G-Protein Coupled Receptor Complexes},
  author={Jones, Roger D and Jones, Alan M},
  journal={Frontiers in Endocrinology},
  volume={14},
  pages={879},
  year={2023},
  publisher={Frontiers}
}

@article{jones2024information,
  title={Information Transmission in G Protein-Coupled Receptors},
  author={Jones, Roger D},
  journal={International Journal of Molecular Sciences},
  volume={25},
  number={3},
  pages={1621},
  year={2024},
  publisher={MDPI}
}

@inproceedings{jones2023proposed,
  title={A Proposed Mechanism for in vivo Programming Transmembrane Receptors},
  author={Jones, Roger D and Jones, Alan M},
  booktitle={Italian Workshop on Artificial Life and Evolutionary Computation},
  pages={123--137},
  year={2023},
  organization={Springer}
}

@article{landauer1996physical,
  title={The physical nature of information},
  author={Landauer, Rolf},
  journal={Physics letters A},
  volume={217},
  number={4-5},
  pages={188--193},
  year={1996},
  publisher={Elsevier}
}

@article{frigg2011entropy,
  title={Entropy-a guide for the perplexed},
  author={Frigg, Roman and Werndl, Charlotte},
  year={2011}
}

@techreport{jones2023vps26,
  title={VPS26 Moonlights as an Arrestin-like Adapter for a 7-transmembrane RGS 2 protein in Arabidopsis thaliana},
  author={Jones, Alan M},
  year={2023},
  institution={University of North Carolina, Chapel Hill, NC (United States)}
}

@article{keshelava2018high,
  title={High capacity in G protein-coupled receptor signaling},
  author={Keshelava, Amiran and Solis, Gonzalo P and Hersch, Micha and Koval, Alexey and Kryuchkov, Mikhail and Bergmann, Sven and Katanaev, Vladimir L},
  journal={Nature communications},
  volume={9},
  number={1},
  pages={1--8},
  year={2018},
  publisher={Nature Publishing Group}
}

@article{latorraca2018molecular,
  title={Molecular mechanism of GPCR-mediated arrestin activation},
  author={Latorraca, Naomi R and Wang, Jason K and Bauer, Brian and Townshend, Raphael JL and Hollingsworth, Scott A and Olivieri, Julia E and Xu, H Eric and Sommer, Martha E and Dror, Ron O},
  journal={Nature},
  volume={557},
  number={7705},
  pages={452--456},
  year={2018},
  publisher={Nature Publishing Group}
}

@article{karshikoff2015rigidity,
  title={Rigidity versus flexibility: the dilemma of understanding protein thermal stability},
  author={Karshikoff, Andrey and Nilsson, Lennart and Ladenstein, Rudolf},
  journal={The FEBS journal},
  volume={282},
  number={20},
  pages={3899--3917},
  year={2015},
  publisher={Wiley Online Library}
}

@article{gerstein2004exploring,
  title={Exploring the range of protein flexibility, from a structural proteomics perspective},
  author={Gerstein, Mark and Echols, Nathaniel},
  journal={Current opinion in chemical biology},
  volume={8},
  number={1},
  pages={14--19},
  year={2004},
  publisher={Elsevier}
}

@article{marsh2012probing,
  title={Probing the diverse landscape of protein flexibility and binding},
  author={Marsh, Joseph A and Teichmann, Sarah A and Forman-Kay, Julie D},
  journal={Current opinion in structural biology},
  volume={22},
  number={5},
  pages={643--650},
  year={2012},
  publisher={Elsevier}
}

@article{jacobs2001protein,
  title={Protein flexibility predictions using graph theory},
  author={Jacobs, Donald J and Rader, Andrew J and Kuhn, Leslie A and Thorpe, Michael F},
  journal={Proteins: Structure, Function, and Bioinformatics},
  volume={44},
  number={2},
  pages={150--165},
  year={2001},
  publisher={Wiley Online Library}
}

@article{sauer2020multi,
  title={Multi-state design of flexible proteins predicts sequences optimal for conformational change},
  author={Sauer, Marion F and Sevy, Alexander M and Crowe Jr, James E and Meiler, Jens},
  journal={PLoS computational biology},
  volume={16},
  number={2},
  pages={e1007339},
  year={2020},
  publisher={Public Library of Science San Francisco, CA USA}
}

@article{bernado2007structural,
  title={Structural characterization of flexible proteins using small-angle X-ray scattering},
  author={Bernad{\'o}, Pau and Mylonas, Efstratios and Petoukhov, Maxim V and Blackledge, Martin and Svergun, Dmitri I},
  journal={Journal of the American Chemical Society},
  volume={129},
  number={17},
  pages={5656--5664},
  year={2007},
  publisher={ACS Publications}
}

@article{teilum2011protein,
  title={Protein stability, flexibility and function},
  author={Teilum, Kaare and Olsen, Johan G and Kragelund, Birthe B},
  journal={Biochimica et Biophysica Acta (BBA)-Proteins and Proteomics},
  volume={1814},
  number={8},
  pages={969--976},
  year={2011},
  publisher={Elsevier}
}

@article{homans2007water,
  title={Water, water everywhere—except where it matters?},
  author={Homans, Steve W},
  journal={Drug discovery today},
  volume={12},
  number={13-14},
  pages={534--539},
  year={2007},
  publisher={Elsevier}
}

@article{teilum2009functional,
  title={Functional aspects of protein flexibility},
  author={Teilum, Kaare and Olsen, Johan G and Kragelund, Birthe B},
  journal={Cellular and Molecular Life Sciences},
  volume={66},
  number={14},
  pages={2231--2247},
  year={2009},
  publisher={Springer}
}

@book{berg2019stryer,
  title={Biochemistry},
  author={Berg Jeremy, M and Tymoczko John, L and Gatto Jr Gregory, J and Stryer Lubert},
  publisher = "W. H. Freeman",
  year={2019}
}

@book{reif2009fundamentals,
  title={Fundamentals of statistical and thermal physics},
  author={Reif, Frederick},
  year={2009},
  publisher={Waveland Press}
}

@article{shannon1963mathematical,
  title={The Mathematical Theory of Communication,(first published in 1949)},
  author={Shannon, CE and Weaver, Warren},
  journal={Urbana University of Illinois Press},
  year={1963}
}

@article{yang2017phosphorylation,
  title={Phosphorylation of G protein-coupled receptors: from the barcode hypothesis to the flute model},
  author={Yang, Zhao and Yang, Fan and Zhang, Daolai and Liu, Zhixin and Lin, Amy and Liu, Chuan and Xiao, Peng and Yu, Xiao and Sun, Jin-Peng},
  journal={Molecular pharmacology},
  volume={92},
  number={3},
  pages={201--210},
  year={2017},
  publisher={ASPET}
}

@article{chen2022qr,
  title={QR code model: a new possibility for GPCR phosphorylation recognition},
  author={Chen, Hao and Zhang, Suli and Zhang, Xi and Liu, Huirong},
  journal={Cell Communication and Signaling},
  volume={20},
  number={1},
  pages={1--16},
  year={2022},
  publisher={Springer}
}

@misc{berg2002biochemistry,
  title={Biochemistry},
  author={Berg, Jeremy M and Tymoczko, John L and Stryer, Lubert and others},
  year={2002},
  publisher={New York: WH Freeman}
}

@book{schrodinger1992life,
  title={What is life?: With mind and matter and autobiographical sketches},
  author={Schrodinger, Roger and Schr{\"o}dinger, Erwin and Dinger, Erwin Schr},
  year={1992},
  publisher={Cambridge university press}
}

@article{phillips2021schrodinger,
  title={Schr{\"o}dinger’s What is life? at 75},
  author={Phillips, Rob},
  journal={Cell systems},
  volume={12},
  number={6},
  pages={465--476},
  year={2021},
  publisher={Elsevier}
}

@article{latorraca2020gpcr,
  title={How GPCR phosphorylation patterns orchestrate arrestin-mediated signaling},
  author={Latorraca, Naomi R and Masureel, Matthieu and Hollingsworth, Scott A and Heydenreich, Franziska M and Suomivuori, Carl-Mikael and Brinton, Connor and Townshend, Raphael JL and Bouvier, Michel and Kobilka, Brian K and Dror, Ron O},
  journal={Cell},
  volume={183},
  number={7},
  pages={1813--1825},
  year={2020},
  publisher={Elsevier}
}

@article{qian2007phosphorylation,
  title={Phosphorylation energy hypothesis: open chemical systems and their biological functions},
  author={Qian, Hong},
  journal={Annu. Rev. Phys. Chem.},
  volume={58},
  pages={113--142},
  year={2007},
  publisher={Annual Reviews}
}
%

%
%\bibliographystyle{splncs04}
%\bibliography{library,acs-achemso}
%

\appendix
\section{Theory}

\subsection{Information and the Second Law of Thermodynamics}
\label{TheoryAppendix}

Standard equilibrium techniques that analyze systems by minimizing free energy do not apply to information-processing systems. Because information processing requires continuous energy and entropy flow, even in steady state \cite{landauer1996physical}, entropy maximization is not achievable in non-isolated systems \cite{jones2025thermodynamics}.

The entropy $S$ of system composed of $W$ states $x\in \mathcal{X}$ is

\begin{equation}\label{entropyAppendix}
    S = - \sum_{i=1}^W \Pr(x_i) \ln \Pr(x_i)
\end{equation}
Assume that the system is composed of a switch in the form of a chemical reaction
\begin{equation}
   \ce{d <=> p}  
\end{equation}
where, as an example, $d$ is the dephosphorylated state of the switch and $p$ is the phosphorylated state of the switch.
The rest of the system is a heat bath that absorbs heat that is generated by the chemical reaction.
The entropy, Eq. \ref{entropyAppendix}, can then be written
\begin{equation}
    S = S_m - I_M
\end{equation}
where $S_m$ is the entropy of the microstates given that the observables $d$ and $p$ is the switch are independent and $I_M$ is the information in the correlation of the mesoscopic components $d$ and $p$ of the switch.
The microscopic entropy $S_m$ is composed of two pieces, the entropy of the microscopic components of the switch and the microscopic entropy of the microstates of the external heat bath.
The entropy of the microstates is
\begin{equation}\label{microstateEntropy}
    S_m =  \sum_{i=1}^W \Pr(x_i,d,p) \ln\left[  \frac{\Pr(d,p)}{ \Pr(x_i) \Pr(d) \Pr(p) }  \right ]
\end{equation}
where $\Pr(d,p)$ is the joint probability of $d$ and $p$.

The mutual information \cite{shannon1963mathematical} associated with the correlation of $d$ and $p$ is
\begin{equation}\label{mutualSwitch}
    I_M = \sum_{\{d,p\} }\Pr(d,p) \ln \left[  \frac{\Pr(d,p)}{\Pr(d) \Pr(p)}  \right]
\end{equation}

The Second Law of Thermodynamics states that the entropy of the system increases with time or achieves a maximum value.
\begin{equation}
    \Delta S = \Delta S_m - \Delta I_M \ge 0
\end{equation}
If the number of states $W$ is infinite, the system never achieves a maximum value.

In the case that the switch achieves a steady state $\Delta I_M \rightarrow 0$, then all the entropy generated by the switch goes into the heat bath.
\begin{equation}
    \Delta S \rightarrow \Delta S_m \ge 0
\end{equation}
When the switch has reached steady state, the switch information $I_M$ is at an extremum.

For a system with an infinite number of states, 
$W\rightarrow \infty$, total entropy continues to increase indefinitely. Local regions of correlation, such as molecular switches, can achieve steady states even as the entropy of the larger system grows. 
The heat generated by the switch is transferred to the surrounding heat bath. This scenario contrasts with finite systems $W$, where the total entropy
$S$ can reach a maximum. Because biological information processing systems function far from equilibrium, traditional frameworks based on minimizing Gibbs free energy $ G = E - T S $, with $E$ as the energy of the system and $T$ the temperature of the bath do not apply.

\subsection{Theoretical Model of the Switch}
\label{Theoretical Model of the Switch}
Building on our earlier theoretical framework \cite{jones14model,jones2023proposed,jones2024information}, which successfully predicted the results of the experimental assay by maximizing the flow of information from the extracellular environment to the intracellular region of the GPCR complex, the present work extends this theory by incorporating a higher temporal resolution. This enhanced model is specifically developed to interpret the detailed, time-dependent dynamics observed in recent impedance-based experiments by Wirth et al. \cite{wirth2023monitoring}. In line with its predecessor, this updated model is guided by the fundamental principle that natural selection favors molecular systems optimized to maximize information transmission capacity \cite{shannon1963mathematical}, thus providing a more granular and dynamic view of the GPCR switching mechanisms.

Figures \ref{SimpleSwitchBody}B and \ref{SimpleSwitchBody}C presents a simplified mathematical representation of the schematic shown in Figure \ref{SimpleSwitchBody}A. This abstraction allows for a more tractable analysis using the tools of statistical mechanics \cite{reif2009fundamentals}.
This level of simplification is common and has been useful in physics \cite{donertacs2011role,ireson2005einstein}. 
In this model, the GPCR complex is reduced to a single switch embedded in a heat bath and surrounded by a flexible protein matrix \cite{karshikoff2015rigidity,gerstein2004exploring,marsh2012probing,jacobs2001protein,sauer2020multi,bernado2007structural,teilum2011protein,teilum2009functional,homans2007water}. The probabilities 
$\Pr(d)$
and 
$\Pr(p)$
represent the fraction of time the receptor spends in the dephosphorylated and phosphorylated states, respectively. Specifically, the
$d$ position corresponds to the state in which the G protein is bound to a GDP molecule, while the 
$p$ position corresponds to the state in which GDP has been phosphorylated to form GTP. 
During a time interval
$\Delta t$, the switch can transition between these states with transition probabilities 
$\Pr(d|p)$ and 
$\Pr(p|d)$, or it can remain in the same state with probabilities 
$\Pr(d|d)$
and 
$\Pr(p|p)$. The heat bath ensures that the fluctuations between these switching states occur as if the system is at a temperature 
$T=1/\beta$,
where 
$T$
is in energy units.

Figure \ref{SimpleSwitchBody}C describes the switch in terms of chemical flux
\begin{equation}
    J_0= \Pr(p|d)\; \Pr(d) = \Pr(d|p)\; \Pr(p)
\end{equation}
which represents the flow of probability from one switch position to another.
This description is useful if the switch is fueled by an external source such as a GTP/GDP disequilibrium.
A flow of energy from GTP/GDP into the switch is turned into heat $Q_{\odot}$ in the time interval $\Delta t$
\begin{equation}\label{entropyFlow}
     Q_{\odot} =  J_0 \; q = -\Delta G_{\odot}
\end{equation}
where $-q$ is the amount of heat released when a GPCR transitions from the ``off" position to the ``on" position and back; and $\Delta G$ is the amount of the free energy from GTP that is injected into the switch in time $\Delta t$.
The minus sign indicates that the heat is leaving the switch.
Equation \ref{entropyFlow} describes the flow of entropy through the switch.
A flow of entropy of this nature is essential for life \cite{morowitz2007energy}.
We take flux $J_0$ as the external driver of emergent behavior in the switching system.

The details of ligand/receptor binding are an important part of the overall picture.
The inclusion of ligand kinetics is shown in Figure \ref{SimpleSwitchBody}D.
The interaction between the ligand and the free receptor
$u$ to form the bound state
$b$ is modeled using Michaelis-Menten kinetics \cite{berg2019stryer}. The forward reaction rate is given by 
$k_+ L$ where $L$ represents the ligand concentration, while the reverse reaction rate is 
$k_-$

Within the bound state
$b$ (Figure \ref{SimpleSwitchBody}), the receptor can adopt different internal states corresponding to the switch positions. The probability that the receptor is in the dephosphorylated ``off-"state
$d$ given that it is in
$b$ is denoted as
$\Pr(d|b)$. 
Similarly, the probability that the receptor is in the phosphorylated ``on" state is
$\Pr(p|a)$.
The observed probabilities $\Pr(d)$ and $\Pr(p)$ are
\begin{equation}
    \Pr(d) = \Pr(b)\; \Pr(d|b)
\end{equation}
and
\begin{equation}
    \Pr(p) = \Pr(b)\; \Pr(p|b)
\end{equation}

Aspects of the ligand/receptor binding kinetics were observed in \cite{wirth2023monitoring}.
Specifically, it was observed that the ligands remained bound to within the time range of the experiment even if the extracellular fluid was washed free of ligands.
The details of these observations provide insight into the role that the switch position plays in ligand binding.
In this study, we do not address this process.
This will be addressed in a follow-up study.

The transmission of information $I$ \cite{shannon1963mathematical,ash2012information,frigg2011entropy} through the switch is: 
\begin{equation}
    I = \sum_{i\in \{ d,p\}}
    \sum_{j\in \{ d,p\}}
    \Pr(j|i)\; \Pr(i)\; \ln \left(  \frac{\Pr(i|j)}{\Pr(i)}  \right)
\end{equation}

\begin{equation}
    I = 
    I(d,p)
    +
    (\Pr(d) - J_0) \ln \left( \frac{\Pr(d|d)}{\Pr(d)} \right)
    +
    (\Pr(p) - J_0) \ln \left( \frac{\Pr(p|p)}{\Pr(p)} \right)
\end{equation}
\begin{equation}
    I(d,p) = J_0 \ln \left( \frac{\Pr(d|p) \; \Pr(p|d)}{\Pr(d)\; \Pr(p)} \right)
\end{equation}
Maximization of the information transmitted through the switch subject to the constraints that the average energy $\bar E$ of the switch is given and the flux $J_0$ is an externally determined constant leads to what we refer to as the Biological Ensemble
\begin{equation*} 
    \beta \Delta E = \beta (E_p-E_d)= 
     \ln
    \left[      
    \frac{Pr(d)^2 \; [\; \Pr(p) - J_0 \;]}{\Pr(p)^2 \;  [ \;\Pr(d)-J_0 \; ]}
    \right]
\end{equation*}
\begin{equation} \label{BiologicalEnsembleCC}
    -\frac{2  (  \Pr(p) -J_0 )}{\Pr(p)}
    +\frac{2  (  \Pr(d) -J_0 )}{\Pr(d)}
    -\frac{2 J_0 [1 - 2 \Pr(p)]}{\Pr(p) \Pr(d)}
\end{equation}

\begin{equation}
    \bar E = \Pr(d)\; E_d + \Pr(p)\; E_p
\end{equation}
where $E_d$ is the energy of the switch in the off position and $E_p$ is the energy of the switch in the on position.
The word ``ensemble" is commonly used in two related concepts.
It can be used to describe the probability given by Eq. \ref{BiologicalEnsembleCC}  of finding a switch in the $p$ position or it can describe an imaginary collection of switch states that have probability given by Eq. \ref{BiologicalEnsembleCC} or another equation such as the Canonical Ensemble \cite{reif2009fundamentals}.

The Biological Ensemble, Eq. \ref{BiologicalEnsembleCC}, determines the theoretical outcomes and testable predictions that can be used to inform the experimental observations of \cite{wirth2023monitoring}.
Like the Canonical Ensemble in equilibrium thermodynamics \cite{reif2009fundamentals}, the Biological Ensemble can predict microscopically observable behavior such as the impedance measurements described in the previous section.
To see this, we assumed that the switch positions are correlated with the GPCR conformations generated by the Z and E forms of the ligand.
It follows that Eq. \ref{BiologicalEnsembleCC} predicts the relationships among the impedance observations in \cite{wirth2023monitoring}

The solutions in Figure \ref{SwitchQuasi} represent the local maximums of information transmission. 
Two parameters determine the system output, the change in internal energy $\beta \Delta E =\beta (E_p-E_d)$ and the externally driven flux $J_0$.
Rapid shifts in these two quantities may initiate a shift from one local maximum to another.

For most pharmaceutical applications, the control parameters are altered by changes in the ligand and ligand conformation.
Of particular importance are jumps initiated by a change in ligand or ligand conformation that move the switch from the ``on position, $\Pr(p) \rightarrow 1$ to the ``off position, $\Pr(d)=1-\Pr(p) \rightarrow 1$.
One type of jump among solutions is particularly important, jumps in which changes in ligand conformation lead to changes in the switch position.
This permits the ligand, which may be a drug, to control the downstream cellular response to the ligand.
In this study, we focus on the identification of these jumps in the observations in \cite{wirth2023monitoring}.
This informs the efficacy of a drug candidate.

\section{Connection Between Experiment and Theory}
\label{Connection Between Experiment and Theory}
The experimenters in \cite{wirth2023monitoring} assumed the existence of a strong correlation between the measured impedance and the downstream response. 
We explicitly assume the simplest relationship between impedance and response, which is linear
\begin{equation}\label{connection}
    \Pr(d) = 1- \Pr(p)=
    \frac{Z-Z_{DMSO}}{Z_{hPP}-Z_{DMSO}},
\end{equation}
where $Z_{hPP}$ is the impedance measured with the endogenous agonist human pancreatic polypeptide (hPP) and $Z_{DMSO}$ is the impedance for the solvent DMSO (negative control).
The Biological Ensemble is symmetric upon interchange of $p$ and $d$.
Therefore, Eq. \ref{connection} with $p$ replaced with $d$ is also a possible connection between experiment and theory.
As long as the relationship between $\Pr(p)$ and $Z$ is monotonic, the conclusions of this paper should not be affected by the exact form of Eq. \ref{connection}. 
hPP and DMSO were used to define the boundaries of observable impedance in \cite{wirth2023monitoring}.
The connection between impedance $Z$ and probability of the ``on" position $\Pr(d)$ provides visibility into how flux $J_0$ and change in free energy $\beta (E_p-E_d)$ change in response to various ligand molecules and conformations of the molecules.
Figure \ref{SwitchQuasi} provides the map for the relationship between impedance and control parameters $J_0$ and change in free energy for each ligand.

\section{Information Landscape}
\label{LandscapeSchematic}
\begin{figure}[hbt!] %s state preferences regarding figure placement here
% use to correct figure counter if necessary
%\renewcommand{\thefigure}{2}
\centering
\includegraphics[width=0.4\textwidth]{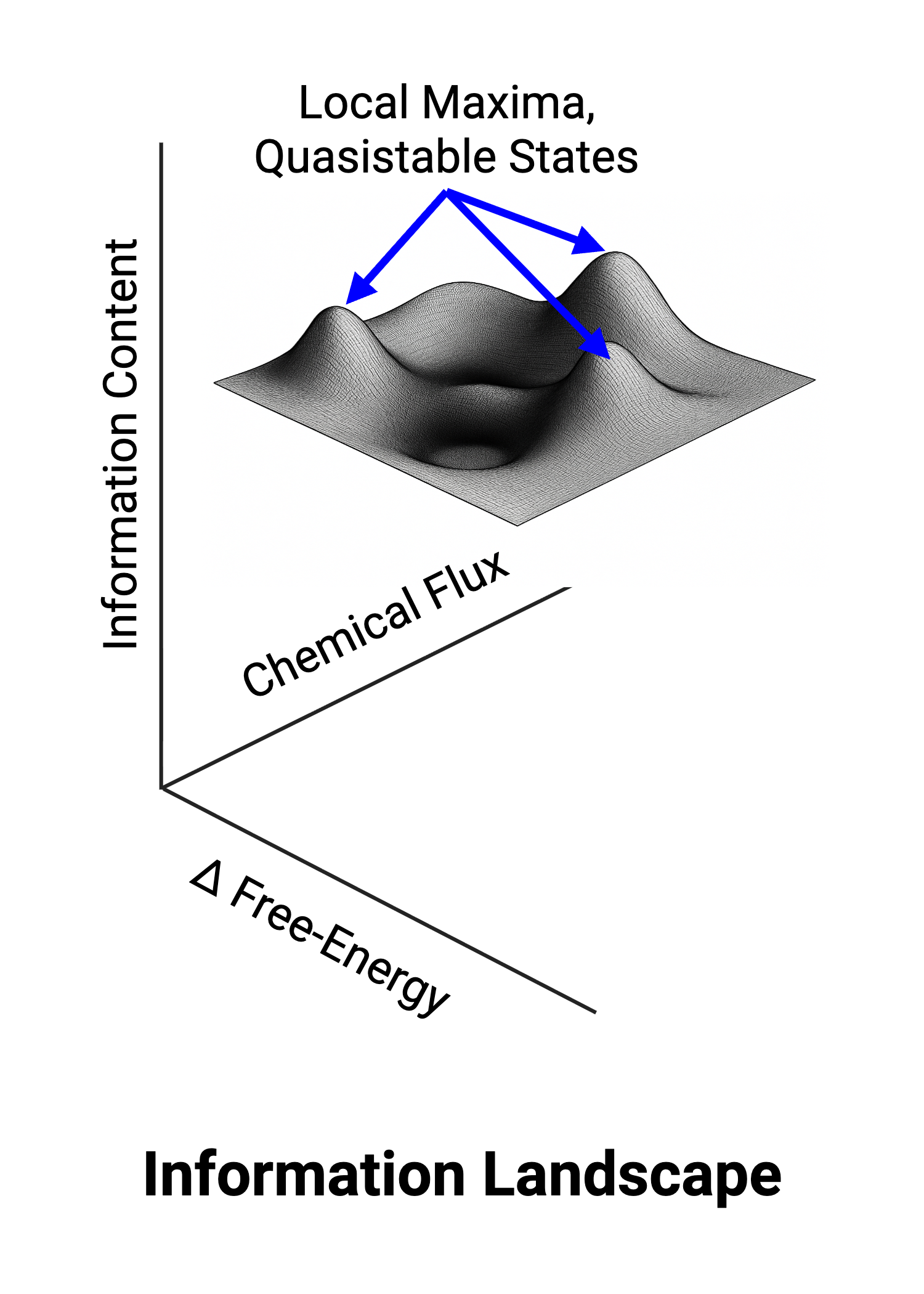}
\caption{ 
The two control parameters are given by the change in the free energy of phosphate bonding, $\beta (E_p -E_d)$, and the chemical flux, $J_0$.
The information content of the system is displayed on the vertical axis.
Local maxima and minima of the information are locations of steady state for information content of the system.
}
\label{Landscape} % \label works only AFTER \caption within figure environment
\end{figure}

Figure \ref{Landscape} presents a conceptual schematic of the information landscape described in Section \ref{TheoreticalResults}, with the two control parameters, $\beta(E_p - E_d)$ and $J_0$, plotted along the horizontal axes and the information content of the system on the vertical axis. 
The surface contains peaks and valleys, representing points where the time derivative of the information content is zero. These extrema correspond to the NESS of the system. 
The specific NESS for the modeled system are mapped in Figure \ref{SwitchQuasi}.
Each ridge in Figure \ref{Landscape} marks a local information maximum, although not necessarily the global maximum.

The local maxima of the Biological Ensemble solutions are shown in Figure \ref{SwitchQuasi}.
For a system residing in one of these local maxima to transition to another local maximum, it must experience a sufficiently large perturbation to overcome the stability of the local state.

At the critical flux
\begin{equation}
    J_0 = \phi_c \approx 0.182
\end{equation}
The maximum information in the thermodynamic branch and the kinetic branch is equal at free energy $\beta \Delta E = \beta (E_p - E_d)$.
As the flux $J_0$ increases , the regimes eventually merge into a single continuous solution at the maximum information value of $J_0=1/2$, as seen in Figure \ref{SwitchQuasi}I.

\end{document}